\documentclass{jfm}

\usepackage{graphicx}
\usepackage{natbib}
\usepackage{color}

\ifCUPmtlplainloaded \else
  \checkfont{eurm10}
  \iffontfound
    \IfFileExists{upmath.sty}
      {\typeout{^^JFound AMS Euler Roman fonts on the system,
                   using the 'upmath' package.^^J}%
       \usepackage{upmath}}
      {\typeout{^^JFound AMS Euler Roman fonts on the system, but you
                   dont seem to have the}%
       \typeout{'upmath' package installed. JFM.cls can take advantage
                 of these fonts,^^Jif you use 'upmath' package.^^J}%
      }
  \else
  \fi
\fi

\ifCUPmtlplainloaded \else
  \checkfont{msam10}
  \iffontfound
    \IfFileExists{amssymb.sty}
      {\typeout{^^JFound AMS Symbol fonts on the system, using the
                'amssymb' package.^^J}%
       \usepackage{amssymb}%
         
       \let\ge=\geqslant  \let\geq=\geqslant
      }{}
  \fi
\fi

\ifCUPmtlplainloaded \else
  \IfFileExists{amsbsy.sty}
    {\typeout{^^JFound the 'amsbsy' package on the system, using it.^^J}%
     \usepackage{amsbsy}}
    {}
\fi

\newsavebox{\astrutbox}
\sbox{\astrutbox}{\rule[-5pt]{0pt}{20pt}}

\title[Closed-Loop Turbulence Control Using Machine Learning]{Closed-Loop Turbulence Control Using Machine Learning}

\author[T. Duriez and friends.]%
{Thomas Duriez$^{1,2}$%
  \thanks{Email address for correspondence: thomas.duriez@gmail.com},\ns
Vladimir Parezanovi{\'c}$^1$,\ns
Laurent Cordier$^1$,\break
Bernd R. Noack$^1$,\ns
Jo{\"e}l Delville$^1$,\ns
Jean-Paul Bonnet$^1$,\break
Marc Segond$^3$\ns
and\ns
Markus Abel$^{3,4,5}$
}

\affiliation{$^1$Institut PPRIME, CNRS - {Universit\a'e de Poitiers} - ENSMA, UPR 3346, D\a'epartement Fluides, Thermique, Combustion, CEAT, 43, rue de l'A\a'erodrome, F-86036 Poitiers Cedex, France\\[\affilskip]
$^2$Laboratoire de M\'ecanique de Lille, Boulevard Paul Langevin,
59655 Villeneuve d'Ascq Cedex, France\\[\affilskip]
$^3$Ambrosys GmbH, Albert-Einstein-Str.\ 1-5, D-14469 Potsdam, Germany\\[\affilskip]
$^4$LEMTA, 2 Avenue de la For\a^et de Haye F-54518 Vandoeuvre-l\a`es-Nancy Cedex, France\\[\affilskip]
$^5$University of Potsdam, Karl-Liebknecht-Str. 24/25 D-14476 Potsdam, Germany}

\pubyear{2014}
\volume{650}
\pagerange{119--126}
\date{?; revised ?; accepted ?. - To be entered by editorial office}
\begin{document}

\maketitle

\begin{abstract}
We propose a general model-free strategy for feedback control design of turbulent flows. This strategy called 'machine learning control' (MLC) is capable of exploiting nonlinear mechanisms in a systematic unsupervised manner. It relies on an evolutionary algorithm that is used to evolve an ensemble of feedback control laws until minimization of a targeted cost function. This methodology can be applied to any non-linear multiple-input multiple-output (MIMO) system to derive an optimal closed-loop control law. MLC is successfully applied to the stabilization of nonlinearly coupled oscillators exhibiting frequency cross-talk, to the maximization of the largest Lyapunov exponent of a forced Lorenz system, and to  the mixing enhancement in an experimental mixing layer flow. We foresee numerous potential applications to most nonlinear MIMO control problems, particularly in experiments.
\end{abstract}

\begin{keywords}
Nonlinear Dynamical Systems/Chaos,
Flow control/Instability control,
Turbulent flows/Turbulence control.
\end{keywords}

\section{Introduction}
Closed-loop turbulence control is a rapidly evolving field of fluid mechanics synergizing many different academic disciplines for engineering applications of epic proportion: drag reduction of transport vehicles, green energy harvesting of wind and water flows, and medical applications, just to name a few.

For many laminar flows, control theory has a well established framework for the stabilization of the steady Navier-Stokes solution based on a local linearization of the Navier-Stokes equation. Corresponding numerical and experimental stabilization studies include virtually any configuration, e.g. wakes ~\citep{Roussopoulos1993jfm}, cavity flows~\citep{Rowley2006arfm,sipp2007jfm,illingworth2012feedback},  
flows of backward-facing step~\citep{Herve2012jfm}, 
boundary-layer flows~\citep{Bagheri2009jfm} 
and channel flow~\citep{Hogberg2003jfm}. 

Turbulent flows pose a number of additional challenges to control design.
First, realistic actuators do not have sufficient authority 
to stabilize the steady Navier-Stokes solution
in contrast to laminar flow.
Second, linear(ized) models cannot resolve 
important frequency cross-talk between the coherent structures, 
the mean flow and the stochastic small-scale fluctuations.
Yet, frequency cross-talk is an important actuation opportunity
as demonstrated by successful wake stabilization with 
high-frequency actuation \citep{Glezer2005aiaaj,Thiria2006jfm,Luchtenburg2009jfm} 
and low-frequency forcing \citep{Pastoor2008jfm}.
Third, model-based control of an experiment 
requires a robust control-oriented reduced-order model 
which is still a large challenge at this moment.
Such a reduced-order model would need --- at minimum --- 
to resolve  the uncontrolled and controlled turbulent coherent structures
including the transients between them.

Experimental studies of closed-loop turbulence control 
are largely based on model-free adaptive approaches.
Most of these experiments start with the finding of an effective periodic actuation.
The actuation amplitude and frequency are slowly adapted 
to maximize an online-monitored performance  \citep{King2010book}.
Prominent examples are extremum seeking for local extrema, 
e.g.\ resonance frequency adaptation, 
and slope seeking for asymptotic convergence, 
e.g.\ amplitude selection \citep{King2006ieeemed}.
These adaptive controls take into account 
the response of all nonlinearities to open-loop forcing,
but they do not provide an in-time response  on time scales of the flow.
One of the few examples of in-time response
is skin friction reduction in wall turbulence.  
Here, a simple opposition control 
in the viscous sublayer is already effective \citep{Choi1994jfm}.
Another example is phasor control 
for turbulence with a dominant oscillatory structure.
In this case, control design requires a robust phase detection from the sensors
and effective gain scheduling for the actuators \citep{Samimy2007jfm}.
A model-based strategy for in-time closed-loop control 
taking into account the relevant nonlinearities is still in its infancy.

In this paper, 
we propose the first model-free alternative
that provides a feedback law to control statistical properties of broadband turbulence. 
Contrary to model-free adaptive control, 
no efficient open-loop control is assumed, 
and the time-scale of the control is the one of the system. 
The methodology, called 'machine learning control' (MLC), 
is based on genetic programming (GP)~\citep{koza1992genetic} and requires only a definition for the objective functional also known as the cost function. 
Genetic programming is a part of the machine learning bundle \citep{wahde2008biologically} 
which has been previously used to design controllers in robotics \citep{lewis1992genetic,nordin1997line}. 
The use of machine learning for control \citep{fleming2002evolutionary} also includes genetic algorithms which can only be used to optimize control parameters \citep{de2014optimizing} and artificial neural networks \citep{noriega1998direct}.

MLC is applied to two simple dynamical systems 
featuring important nonlinearities of turbulence.
The first plant is a generalized mean-field model 
with two nonlinearly coupled oscillating constituents, for which controllers based on the linearized model fail. 
The second one is a forced Lorenz system 
for which we demonstrate the original use of the cost function: 
we want to maximize chaos. 
Its applications may lie in mixing systems like in combustion.
The main demonstration of MLC is an experimental mixing enhancement in the TUCOROM mixing layer wind-tunnel.

The manuscript is organized as follows:
In \S \ref{ToC:MLC}, the machine learning control strategy is described.
The three chosen control problems and associated cost functions 
are defined in \S \ref{ToC:plants}
followed by the corresponding results in \S \ref{ToC:results}.
Conclusions and future directions are provided in section \S \ref{ToC:conclusion}.

\section{Machine learning control}\label{ToC:MLC}
In the following, we restrict the description to ordinary differential equations for reasons of comprehensibility. The system is represented in phase space by the vector $\mathbf{a}\in \mathbb{R}^{n_a}$, it is measured by sensors $\mathbf{s}\in \mathbb{R}^{n_s}$, and controlled  by actuators $\mathbf{b}\in \mathbb{R}^{n_b}$,
\begin{equation}
\frac{d\mathbf{a}}{dt} = \mathbf{F} \left( \mathbf{a}, \mathbf{b}\right)\, ,\;
\mathbf{s}       = \mathbf{H} \left( \mathbf{a}\right)\, ,\;
\mathbf{b}  = \mathbf{K} \left(\mathbf{s}\right) \,,
\end{equation}
with $\mathbf{F}$ denoting a general nonlinear function, $\mathbf{H}$ the measurement function, and $\mathbf{K}$ the sensor-based control law. This law shall minimize the state- and actuation-dependent cost function:
\begin{equation}
J = J(\mathbf{a},\mathbf{b}).
\end{equation}
The cost function value grades how a given control law $\mathbf{K}(\mathbf{s})$ performs relatively to the problem at stake. The lower the value of the cost function, the better the control law solves the problem.

\begin{figure}
\centerline{
\begin{tabular}{ccc}
\includegraphics[width=0.4\textwidth]{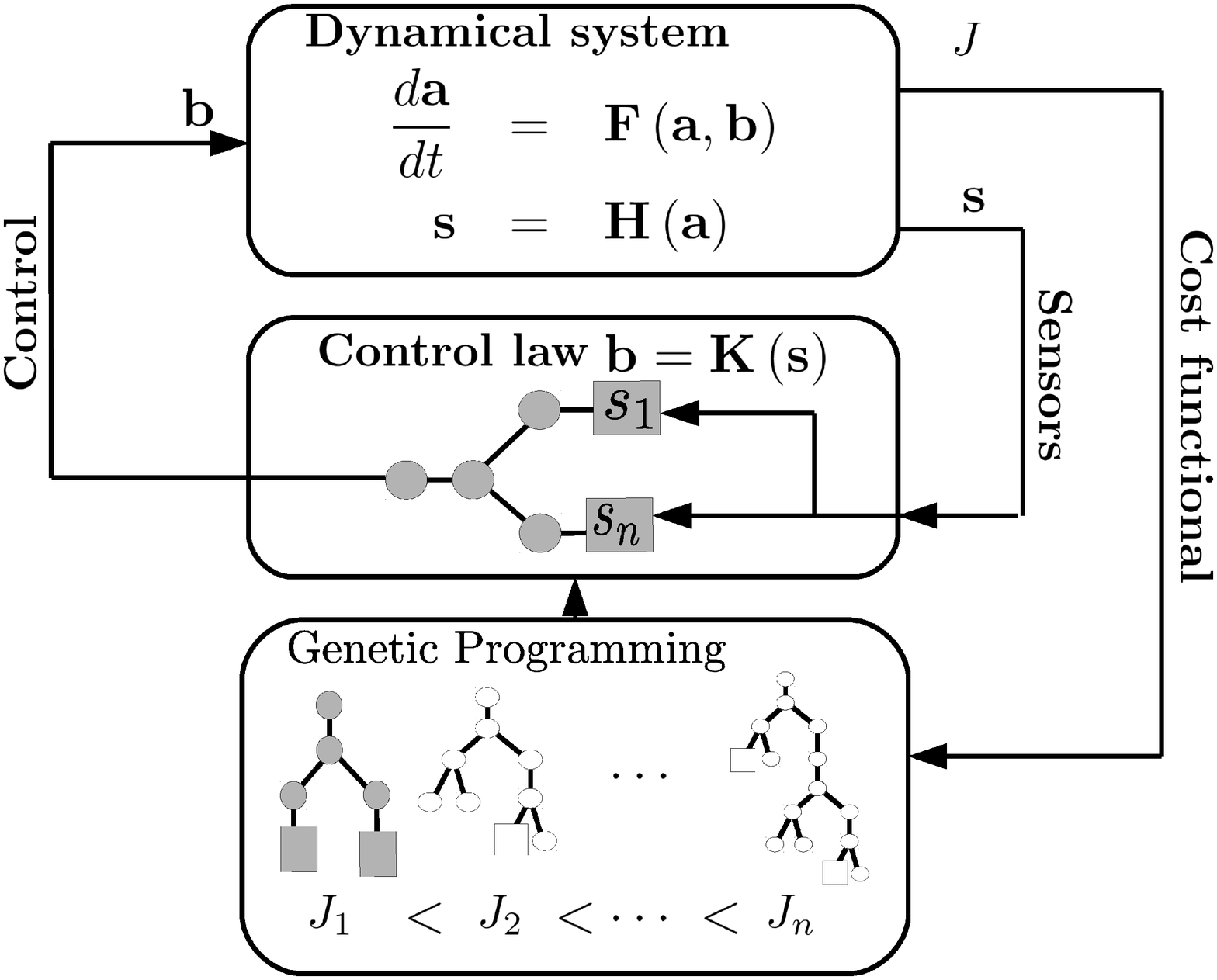}&\qquad\qquad &
\includegraphics[width=0.4\textwidth]{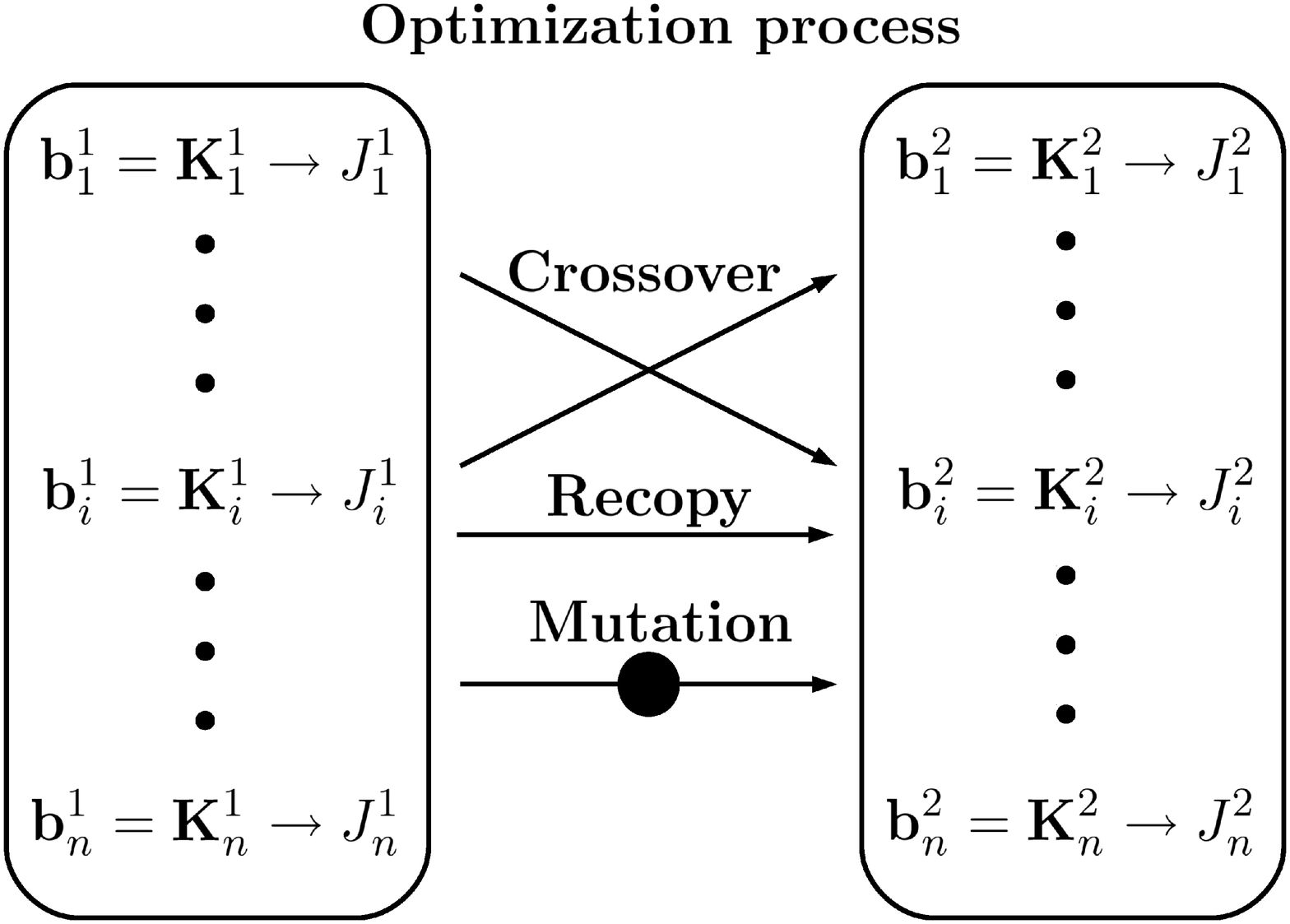}
\end{tabular}
}
\caption{Left: Control design using MLC. During a learning phase, each control law candidate is evaluated by the dynamical system or experimental plant. This process is iterated over many generations of individuals. At convergence, the best individual (in grey) is determined and used for control. Right: Production of a new generation of individuals. Each individual $K_{i}^{m}$ is ranked by their cost, $J_{i}^{m}$, $i$ pointing to the $i^{th}$ individual, $m$ to the $m^{th}$ generation. An individual of the subsequent
generation can be a copy, a mutation or the result of the cross-over of individuals selected in the preceding generation according to their cost.
} 
\label{fig:GPprocess}
\end{figure}

We propose a model-free design of the control law. The genetic programming is used to design the best control law $\mathbf{K(s)}$ as a composition of elementary functions. A first set of control law candidates (called individuals) is generated by random compositions of selected elementary functions. 
The employed GP algorithm~\citep{ecj} combines these operations as a tree \citep{koza1992genetic}
to generate effectively any linear or nonlinear function. Each individual is attributed a cost via the evaluation of $J(\mathbf{a},\mathbf{b})$. The next set  of individuals (called generation) is generated by mutation, cross-over or replication of individuals with a specific rate for each process (see figure~\ref{fig:GPprocess}). 

The individuals used to produce the new generation are selected based on how well they minimize the cost function. A global extremum of the cost function is typically approximated well in a finite number of generations if the population contains enough diversity to explore the search space. The method has been shown to be successful~\citep{lewis1992genetic,nordin1997line} even though there is no general mathematical proof for convergence.

\section{Control problems}\label{ToC:plants}
MLC is used to put the system in a desirable state as equilibrium (\S \ref{ToC:desGMM}), to optimize a given measure on the system such as Lyapunov exponents (\S \ref{ToC:desLor}) or the width of a turbulent mixing layer (\S \ref{ToC:desExp}).
\subsection{Generalized mean-field model}\label{ToC:desGMM}
We first consider a generalized mean-field model describing frequency cross-talk for a variety of physical phenomena including fluid flows~\citep{zielinska1997strongly,Luchtenburg2009jfm}. This model can be viewed as a generalization of the Landau model for the bifurcation from equilibrium to a periodic oscillation. Since we focus on frequency cross-talk, we choose a simple form of this model with two oscillators coupled by a nonlinear variation of one growth rate:
\begin{eqnarray}
\frac{\mathrm{d}}{\mathrm{d}t}\left[
\begin{array}{c}
a_1\\
a_2\\
a_3\\
a_4
\end{array}\right]
= \left[\begin{array}{cccc}
\sigma_1  & \omega_1 & 0 & 0 \\
-\omega_1 & \sigma_1 & 0 & 0 \\
0 & 0 & \sigma_2  & \omega_2 \\
0 & 0 & -\omega_2 & \sigma_2
\end{array}\right]
\left[
\begin{array}{c}
{a_1}\\
{a_2}\\
{a_3}\\
{a_4}
\end{array}\right]
+
\left[
\begin{array}{c}
0\\
0\\
0\\
b
\end{array}\right]\label{eq:GMM1}\\
\mbox{with }\sigma_1 = \sigma_{10} - (a_1^2 +a_2^2 + a_3^2 +a_4^2).\nonumber
\end{eqnarray}
Hereafter, we denote the sum of squared amplitudes as energy to avoid linguistic sophistication. We set $\omega_1=\omega_2/10=1$ and $\sigma_{10}=-\sigma_2=0.1$ so that the first oscillator $(a_1,a_2)$ is unstable at the origin while the other one $(a_3,a_4)$ is stable. When uncontrolled ($b\equiv 0$), the nonlinearity drives the first oscillator to nonlinear saturation by the change of total energy. The actuation  directly effects only the stable oscillator. This  system is arguably the simplest nonlinear dynamical system to exhibit frequency cross-talk. We choose to stabilize the first oscillator around its fixed point $(0,0)$ and thus a cost function which measures the fluctuation energy of that unstable oscillator. For any useful application, the energy used for control is required to be small, hence, we penalize the actuation energy: 
\begin{equation}
J=\left< a_1^2(t) + a_2^2(t) + \gamma b^2(t) \right>_T,
\label{eq:fitness}
\end{equation}
with $\gamma = 0.01$ as penalization coefficient and $\left<\cdot\right>_T$ denoting the average over the time interval $[0, T]$. Here, $T=100\times{2\pi}/{\omega_1}$ is chosen to allow meaningful statistics. The quadratic form of the state and the actuation in the cost function is a standard choice in control theory. We apply MLC with full-state observation ($\mathbf{s}\equiv\mathbf{a}$) to exploit all potential nonlinear mechanisms to control the unstable oscillator.

Knowing the nonlinearity at stake, an open-loop strategy can be designed: exciting the stable oscillator at frequency $\omega_2$ will provoke an energy growth which stabilizes the first oscillator as soon as $a_1^2 +a_2^2+a_3^2 +a_4^2 > \sigma_{10}$. Note that the linearization of (\ref{eq:GMM1}) yields two uncoupled oscillators. Thus, the first oscillator is uncontrollable in a linear framework. 

\subsection{Lorenz system}\label{ToC:desLor}
As second example, we consider the Lorenz system controlled in the third component:
\begin{eqnarray}
\frac{\mathrm{d}a_1}{\mathrm{d}t} &=&  \sigma\left(a_2 - a_1\right),\nonumber\\
\frac{\mathrm{d}a_2}{\mathrm{d}t} &=& a_1\left(\rho-a_3\right)-a_2,\\
\frac{\mathrm{d}a_3}{\mathrm{d}t} &=&  a_1a_2 - \beta a_3 +b,\nonumber
\label{eq:Lorenz} 
\end{eqnarray}
with full-state feedback $b=K(a_1, a_2, a_3)$, i.e.\ $\mathbf{s}\equiv\mathbf{a}$. The Lorenz system can be stable, periodic or chaotic depending on the set of used parameters. We employ $\sigma=10$, $\beta=8/3$ and $\rho=20$, such that the uncontrolled system ($b\equiv0$) is periodic. Instead of stabilizing an equilibrium, we demonstrate how to obtain a chaotic system. Existing strategies may stabilize or destabilize periodic orbits~\citep{ott1990controlling,pyragas1992continuous,schusterhandbook}. Like \cite{de2014optimizing},  we aim at maximizing the largest Lyapunov exponent $\lambda_{1}$ while penalizing the actuation power with a factor $\gamma$. If $\lambda_{1}$ is positive, the system is chaotic and well-mixing. We define the cost function, which should be minimized, as:

\begin{equation}
\begin{array}{l l l l}
J   & =           & \exp(-\lambda_{1}) +\gamma\left< b^2(t)\right>_T  &\quad\mbox{if}\,\, \sum_{i=1}^3 \lambda_i < 0,\\
J  &\rightarrow & \infty&\quad\mbox{if}\,\, \sum_{i=1}^3 \lambda_i \geq 0,
\end{array}
\end{equation}
where $T=100$ is the integration time and $\lambda_1\ge\lambda_2\ge\lambda_3$ are the Lyapunov exponents. 
These exponents are obtained by a standard algorithm~\citep{wolf1985determining}. 
$J$ is assigned the largest computable real number on the computer 
if the sum of the Lyapunov exponents is positive or the states exceed the bounds we specify. 

\subsection{Experimental mixing layer}\label{ToC:desExp}
\begin{figure}
\begin{center}
\includegraphics[width=0.8\textwidth]{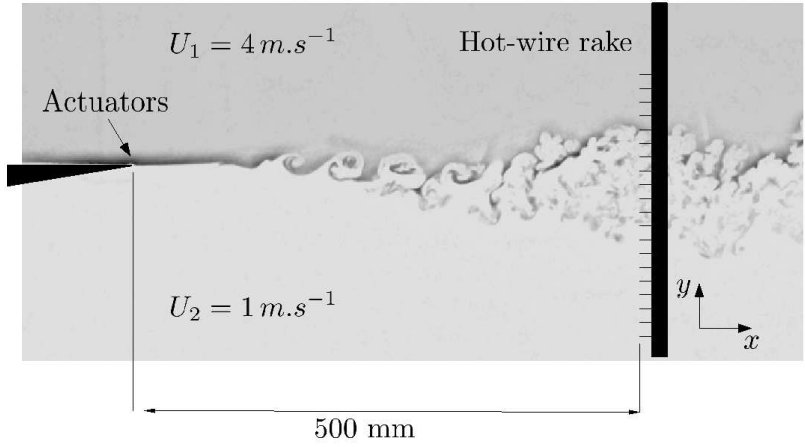}
\end{center}
\caption{Experimental setup of the mixing layer. 
The hot-wire rake is placed at $500\,mm$ downstream of separating plate 
to capture the structures in the shear layer. 
The spacing of the hot-wire probe is $\delta{y}=8\,mm$.\label{fig:expsetup}}
\end{figure}

The TUCOROM mixing layer experimental demonstrator is a dual stream wind tunnel with independently controlled turbines. The test section after the trailing edge of the separating plate is of dimension $width\times height \times length = 1.0 \times 1.0 \times 3.0 \,m^3$. In this experiment, the Reynolds number is $Re_\theta=U_c\theta/\nu= 500$ based on convective velocity $U_c=(U_1+U_2)/2$ and shear layer momentum thickness $\theta$. The sensors are made of a rake of 24 hot-wires to record velocity fluctuations on a vertical profile across the shear layer. The actuators are 96 micro jets located at the tip of the separating plate and which can be triggered up to $500\,Hz$ (see figure~\ref{fig:expsetup}). The machine learning control strategy is applied to maximize the width of the mixing layer,  
\begin{equation}
J=\frac{1}{W},\,\, \mbox{with}\,\, W=\frac{\left<\left[\sum_{i=1}^{24}{s}_i^2(t)\right]\right>_T}{\mbox{max}_{i \in [1,24]}(\left<{s}_i^2\right>_T)},\label{eq:Jexp}
\end{equation}
where ${s}_i(t)$ is the velocity fluctuation as recorded by the hot-wire anemometer $i$, $\left<\cdot\right>_T$ is an average of all acquisitions during the evaluation time $T=10\,s$ corresponding to about 1000 Kelvin-Helmholtz period. This cost function is minimized when the width $W$ of the fluctuation energy profile is maximized. 

\section{Results}\label{ToC:results}
In this section, we present the results of the MLC algorithm for the three examples discussed in \S \ref{ToC:plants}: a system with frequency cross-talk (\S \ref{ToC:resGMM}), an optimization of chaos (\S \ref{ToC:resLor}) and an experiment with a turbulent mixing layer (\S \ref{ToC:resChex}).
\subsection{Generalized mean-field model}\label{ToC:resGMM}
The function space is explored by using a set of elementary ($+,-,\times,/$) and transcendental (e.g. $\exp,\sin,\ln$) functions. The functions are 'protected' to allow them to take arbitrary arguments in $\mathbb{R}$. Additionally, the actuation command is limited to the range $[-1\,,\, 1]$ to emulate an experimental actuator. Up to $50$ generations comprising $1000$ individuals  are processed. 

\begin{figure}
\centerline{\includegraphics[width=0.8\textwidth]{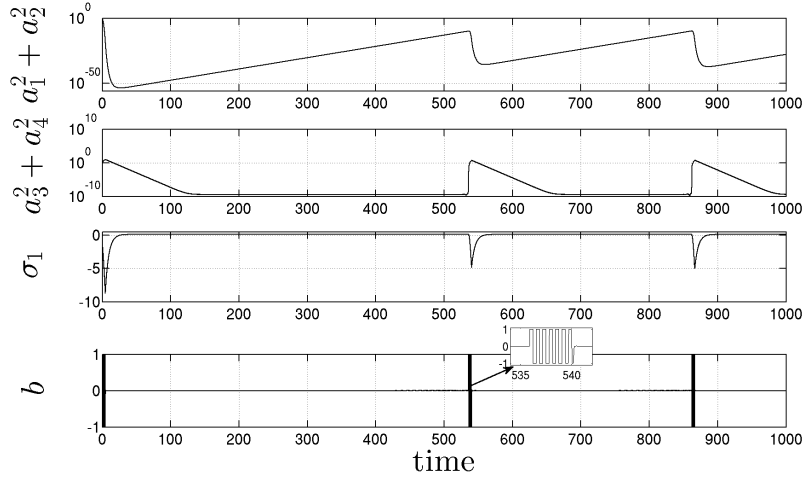}}
\caption{Controlled generalized mean-field model. When the energy contained in the first oscillator (top) is larger than $10^{-10}$ the control (bottom) is exciting the second oscillator at frequency $\omega_2$, its energy grows so that $\sigma_1$ reaches approximately $-5$. This results in a fast decay of the energy in the first oscillator after which the control goes in ``standby mode``. An animation of the controlled system can be found in \S \ref{ToC:sup}.}
\label{fig:energies}
\end{figure}

The control law ultimately returned by the MLC process corresponds to the best individual of the last generation.  The formula is given in \S \ref{ToC:sup}. It can be summarized as follows:
\begin{equation}
b=K_1(a_4)\times K_2(a_1,a_2,a_3,a_4).
\end{equation}
The function $K_1(a_4)$ describes a phasor control that destabilizes the stable oscillator. The function $K_2(a_1,a_2,a_3,a_4)$ acts as a gain dominated by the energy of the unstable oscillator. The performance and the behaviour of the control law are displayed in figure~\ref{fig:energies}. The control law is energizing the second oscillator up to $10^0 \gg \sigma_{10}$ as soon as the first oscillator has an energy which is larger than $10^{-10}$. This is stabilizing the unstable oscillator very quickly with a decay scaling roughly as $\exp(-10^5t)$. After stabilization, the control stays at very low values. That keeps the stable oscillator at a correspondingly low energy $\approx 10^{-10}$, 
while the amplitude of the unstable oscillator is exponentially increasing with its initial growth rate $\sigma_{10}$. This control law exploits the frequency cross-talk and vanishes when not needed, i.e. $a_1\approx a_2\approx 0$. That control could not be derived from a linearized model of the system. Less energy is used as compared to the best periodic excitation.

\subsection{Lorenz system}\label{ToC:resLor}
\begin{figure}
\begin{center}
\begin{tabular}{c}
\centerline{\includegraphics[width=0.9\textwidth]{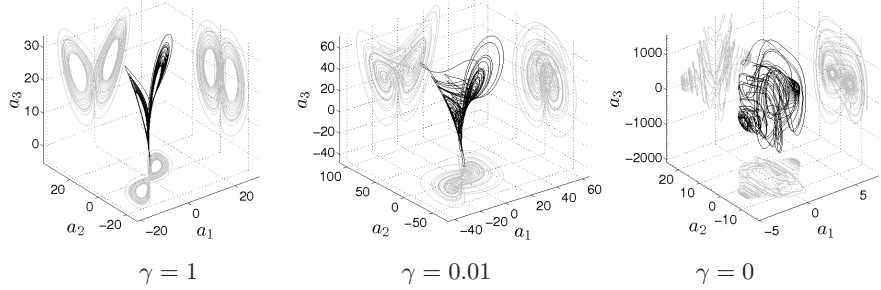}}\\
\hfill $\gamma=1$ \hfill $\gamma=0.01$ \hfill $\gamma=0$ \hfill \mbox{}
\end{tabular}
\end{center}
\caption{Controlled Lorenz systems with $\sigma=10$, $\beta=8/3$ and $\rho=20$. For $\gamma=1$ (left), the system exhibits chaotic behaviour ($\lambda_{1}=0.715$) close to the canonical chaotic Lorenz attractor with $\rho=28$ ($\lambda_{1}=0.905$). For $\gamma=0.01$ (center), the system exhibits more complex trajectories, the nature of the central fixed point has changed and $\lambda_{1}=2.072$. For $\gamma=0$ (right), the nature of all fixed points has changed. The non-penalization of the actuation leads to a change in the scales ($\lambda_{1}=17.613$). An animation of the controlled system can be found in \S \ref{ToC:sup}.}
\label{fig:3DLorenz}
\end{figure}

MLC is applied to the periodic Lorenz system to maximize the largest Lyapunov exponent while keeping the solution bounded. The basic operations that compose the control law are ($+$, $-$, $\times$, $/$) as well as randomly generated constants. The maximum number of generations is chosen as 50 with 1000 individuals each. 
We consider for $\gamma$ the values of $\gamma_S=1$, $\gamma_W=0.01$ and $\gamma_N=0$, representing strong, weak and no penalization of the actuation.
This illustrates how the cost function definition influences the problem to be solved. 
After 50 generations, the best individuals (see \S \ref{ToC:sup}) associated with strong, weak and no penalization have maximum Lyapunov exponents of $\lambda_{1}=0.715$, $2.072$ and $17.613$, respectively. The changes in the system and the control function are displayed in figure~\ref{fig:3DLorenz}. The control laws associated with $\gamma_S$ and $\gamma_W$ are affine expressions of $a_3$ and the reduction of the actuation cost leads to a larger amplitude of the feedback. In those cases, the most efficient controls lead the system into behaviours close to the canonical Lorenz system ($\rho=28$, $\lambda_{1}=0.905$). For $\gamma_W$ the nature (from saddle point to spiral saddle point) and the position of the central fixed point from the actuated system are changed. If the actuation is not penalized ($\gamma=0$) the feedback law is a complex nonlinear function of all states. The nature and position of all fixed points are changed as $\lambda_{1}$ reaches higher values. 


\subsection{Experimental mixing layer}\label{ToC:resChex}
\begin{figure}
\begin{center}
\includegraphics[width=0.9\textwidth]{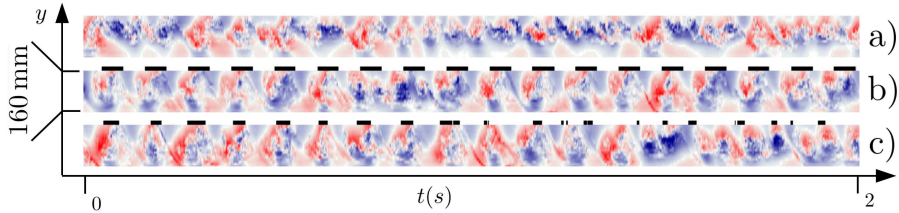}
\end{center}
\caption{Pseudo-visualizations of the TUCOROM experimental mixing layer demonstrator~\citep{Parezanovic2013tsfp}
for three cases:
(a) unforced baseline (width $W=100$\%),  
(b) the best open-loop benchmark (width $W=155$\%)
and (c) MLC closed-loop control (width $W=167$\%).
The velocity fluctuations recorded  by 24 hot-wires probes (see figure \protect\ref{fig:expsetup})
are shown as contour-plot over the time $t$ (abscissa) 
and the sensor position $y$ (ordinate).
The black stripes above the controlled cases indicate 
when the actuator is active (taking into account the convective time). 
The average actuation frequency achieved by the MLC control is comparable to the open-loop benchmark.}
\label{fig:chex}
\end{figure}

MLC is applied to the TUCOROM experimental mixing layer demonstrator~\citep{Parezanovic2013tsfp}. The selected functions are $(+,-,\times,/,\sin,\cos,\log,\tanh)$. The micro-jets are turned on if the action command is positive and off otherwise. 
The number of generations is chosen to be 25 with 100 individuals each. The evaluation of one generation is achieved in 40 minutes of experiment. The employed cost function (\ref{eq:Jexp}) maximizes the width of the fluctuation energy profile in the shear layer. The control law ultimately returned by the MLC algorithm is compared to the reference open-loop control, an harmonic forcing at the most efficient frequency as determined by a parametric study (see figure~\ref{fig:chex}). While the best open-loop forcing is able to upgrade the cost function value by 55\% (compared to the uncontrolled flow), MLC is able to improve it by 67\%. Moreover, the total flow-rate through the actuation jets achieved by the MLC closed-loop control is reduced by 46\%. These results shall be further detailed in an upcoming publication.

\section{Conclusions and future directions}\label{ToC:conclusion}
We propose a model-free optimization of sensor-based control laws 
for general multiple-input multiple-output (MIMO) plants, 
the 'machine learning control' (MLC).
This strategy is based on genetic programming (GP).
GP is one of the most versatile methods for function optimization in machine learning
and includes genetic algorithms (GA).
While GA performs a parameter identification of a given control law, 
GP performs also a structure identification of arbitrary nonlinear control laws.
Thus, MLC comprises an increasingly popular control optimization based on GA. 
MLC is based on an ensemble (called 'generation')
of general nonlinear functions (called 'individuals')
and invests in an exploration of novel laws.
Thus, MLC has a large chance to detect and exploit otherwise invisible local extrema.
In contrast, 
model-free adaptive control is particularly suited 
for adjusting one or few parameters 
of prescribed open- or closed-loop control laws to changing flow conditions.
Such online parameter adaptation
is not part of the presented MLC method 
but could --- in principle --- be included.

As our first test-case,
MLC has been successfully applied 
to a closed-loop stabilization of a generalized two-frequency mean-field model 
detecting and exploiting frequency cross-talk in an unsupervised manner. 
Frequency cross-talk is of primordial importance 
for large-scale turbulence control
with complex interactions between the coherent structures at different dominant frequencies, 
the mean flow changing on  large time scales  
and the cascade to small-scale structures with small associated time scales.
By definition, frequency cross-talk is ignored in any linearized system. Another successful demonstration of MLC 
is closed-loop control for the maximization of
the Lyapunov exponent (stretching) 
of the forced Lorenz equations.
Again, this increase of unpredictability is a highly nonlinear phenomenon.

A challenging experimental closed-loop control demonstration 
is the increase of the mixing layer width in the TUCOROM wind-tunnel \citep{Parezanovic2013tsfp}.
MLC outperforms the best periodic forcing by an additional 12\% increase of the mixing layer width 
and leads to a  significant reduction of the actuation cost.
It may be noted that this open-loop reference level 
is also obtained with an extremum seeking method.
Expectantly, corresponding adaptive control does not lead to a better mixing.
MLC has overcome important technical challenges for in-time control:
(1) the hot-wire sensors show broadband frequency dynamics,
(2) the large convective time delay from actuators to sensors and 
(3) this response was found to be strongly nonlinear.

Summarizing, the model-free formulation of MLC gives rise to a high flexibility: 
it can be applied to any MIMO plant and use any cost function.
Though a model is not needed, 
the more we know about the system, 
the better we can design the cost function according to the underlying physics
and the better we can bias the control law selection.
Further improvements can be expected from including actuation or sensor histories,
like in ARMAX models \citep{Herve2012jfm}.
The relation of tree depth, number of generations, number of individuals with convergence is subject of ongoing research and may boost the performance considerably. 

The major drawback of the model-free approach lies in the evaluation time, 
as each individual needs a simulation or experiment to be run. 
This translates in a large time requirement should the process be serial. 
Consequently, massive parallelization of computations or experiments 
may be needed in real-world MLC applications.
For instance, transition control in a pipe flow may be performed
with a grid of 10 times 10 simultaneously used parallel pipes.

The model-free control design is particularly interesting 
for experimental applications for which a model might not even be known,
like for the control of some multi-phase or multi-physics flows 
with several phases, combustion or unknown non-Newtonian fluids. 
We conjecture that MLC will play a similar role as control theory
in the closed-loop control of turbulence and other complex flows.

\section{Acknowledgements}
We acknowledge funding of the French Science Foundation ANR 
(Chaire d'Excellence TUCOROM, SepaCoDe). MS and MA acknowledge the support of the LINC project (no. 289447) funded by EC’s Marie-Curie ITN program (FP7-PEOPLE-2011-ITN).
We thank Steven Brunton, Eurika Kaiser and Nathan Kutz for fruitful discussions and comments.

\section{Supplementary material}\label{ToC:sup}
\noindent
A document displaying the control laws derived by MLC is available as supplementary material. A movie displaying the behaviour of the controlled generalized mean-field model is available as Movie 1. A movie displaying the behaviour of the controlled Lorenz system is available as Movie 2

\bibliographystyle{jfm}


\end{document}